\documentclass[10pt,letterpaper]{article}
\usepackage{opex3}
\usepackage{color}
\usepackage{epstopdf}

\begin{document}

\title{Local density of states analysis of surface wave modes on truncated photonic crystal surfaces with nonlinear material}

\date{\today}
\author{J. Merle Elson and Klaus Halterman}

\address{Naval Air Warfare Center, Research Department, Signal and Sensor Sciences Division, China Lake, CA 93555}

\email{john.elson@navy.mil} 



\begin{abstract}
The local density of states and response to an incident plane wave of a finite sized photonic crystal (PC) with nonlinear material (NLM) is analyzed. Of particular interest is the excitation of surface wave modes at the truncated surface of the PC, which is collocated with the NLM material. We compute the 2D Green function of the PC with linear material and then include the Kerr NLM in a self-consistent manner. The 2D PC consists of a square array of circular rods where one row of the rods is semi-circular in order to move the surface wave defect mode frequency into the band gap. Since the surface modes are resonant at the interface, the NLM should experience at least an order of magnitude increase in field intensity. This is a possible means of increasing the efficiency of the PC as a frequency conversion device.
\end{abstract}




\section{Introduction}
The refinement of lithographic and other fabrication techniques used in photonic crystal (PC) development has generated a broad range of experimental, and theoretical interest \cite{yablonovitch,robertson,hillebrand,temelk,konoplev,wang,fussell}. In particular, the ability of PC structures to 
control the propagation of electromagnetic fields has stimulated certain emerging concepts and devices that are
based on PC technology, including microscopic lasers, optical switches, light localization, and resonant cavity antennas \cite{temelk}.

It is computationally convenient when investigating PC structures, to assume the PC is of infinite extent.  
This means a perfect crystal, or a crystal with defects in a supercell configuration, and this works well for band structure calculations. This approximation has limitations, since experimental situations involving finite-sized crystals may involve defect modes that result from truncating the infinite periodicity, or reflections and interference effects of the fields around the boundaries \cite{robertson}. The vast  majority of theoretical studies also assume the PC consists of linear material. When characterizing the electromagnetic  properties of PC structures,
the transmittance characteristics often yield limited information regarding the band gap, owing mainly to the directional dependence of the incident field \cite{wang}. The quantity of experimental relevance, that affords a more 
complete and fundamental picture of the spectra and modes of the system, is the spatially and energy resolved local density of states (LDOS).\cite{fussell}

In this paper, we compute LDOS and field characteristics for a finite two-dimensional PC consisting of both linear and nonlinear material. Some researchers \cite{villa} have focused on surface modes associated with a one-dimensional PC. Our emphasis is on the properties of surface resonant modes as a localized field to concentrate the electric field in the presence of NLM. For the case of two-dimensional truncated structures consisting of an array of cylinders,  surface electromagnetic waves can be supported. These surface modes are localized near the truncation boundary 
and propagate parallel to the interface.  Since the intensity of the surface  modes is concentrated at the interface, and if the interface contains nonlinear material (NLM), then a more efficient frequency conversion might be achieved. This could be useful in frequency conversion devices for use in detectors to operate between visible and infrared wavelengths.

In Sec.\ref{theory}, we outline the integral formulation of the problem. Next, we briefly discuss the iterative numerical algorithm used to calculate the electromagnetic fields and Green's function, which gives essential information regarding the LDOS. The LDOS will help reveal certain defect modes of the PC that occur at frequencies within a band gap. We present the results in Sec.\ref{results} and find that within the PC structure, the LDOS is small but nonzero within the band gap, and that any localized states reside mainly within the truncated surface. Results are also shown that illustrate the effect of incorporating NLM material.

\section{Theory \label{theory}}
\begin{figure}
\centerline{\includegraphics[width=6cm]{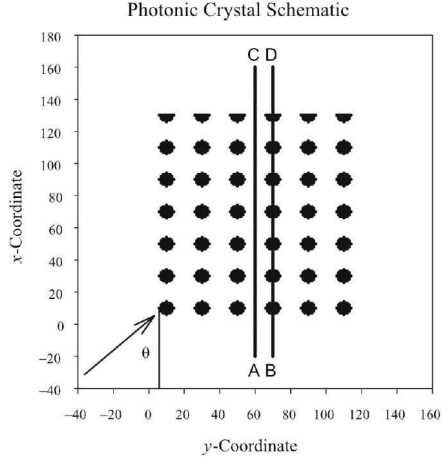}}
\caption{{\label{pc}}Schematic of photonic crystal (PC) cross-section. The PC consists of infinitely long rods in a 2D array of 7 rods $\times$ 6 rods. Note that the rods in the topmost layer  have been truncated along their diameter into a semi-circular shape. This truncation will allow localized surface modes to exist along this side of the PC within band gap frequencies. The lines AC and BD are guides for the eye in relating to numerical results below. The arrow and angle $\theta$ denote an incident plane wave.  The PC has period $a$, the cylinders have radius $r = 0.2a$ with permittivity $\epsilon_c = 8.9$, and the background medium is vacuum. Dispersion characteristics of this PC are given in Fig. \ref{band}. }
\end{figure}

\begin{figure}
\centerline{\includegraphics[width=6cm]{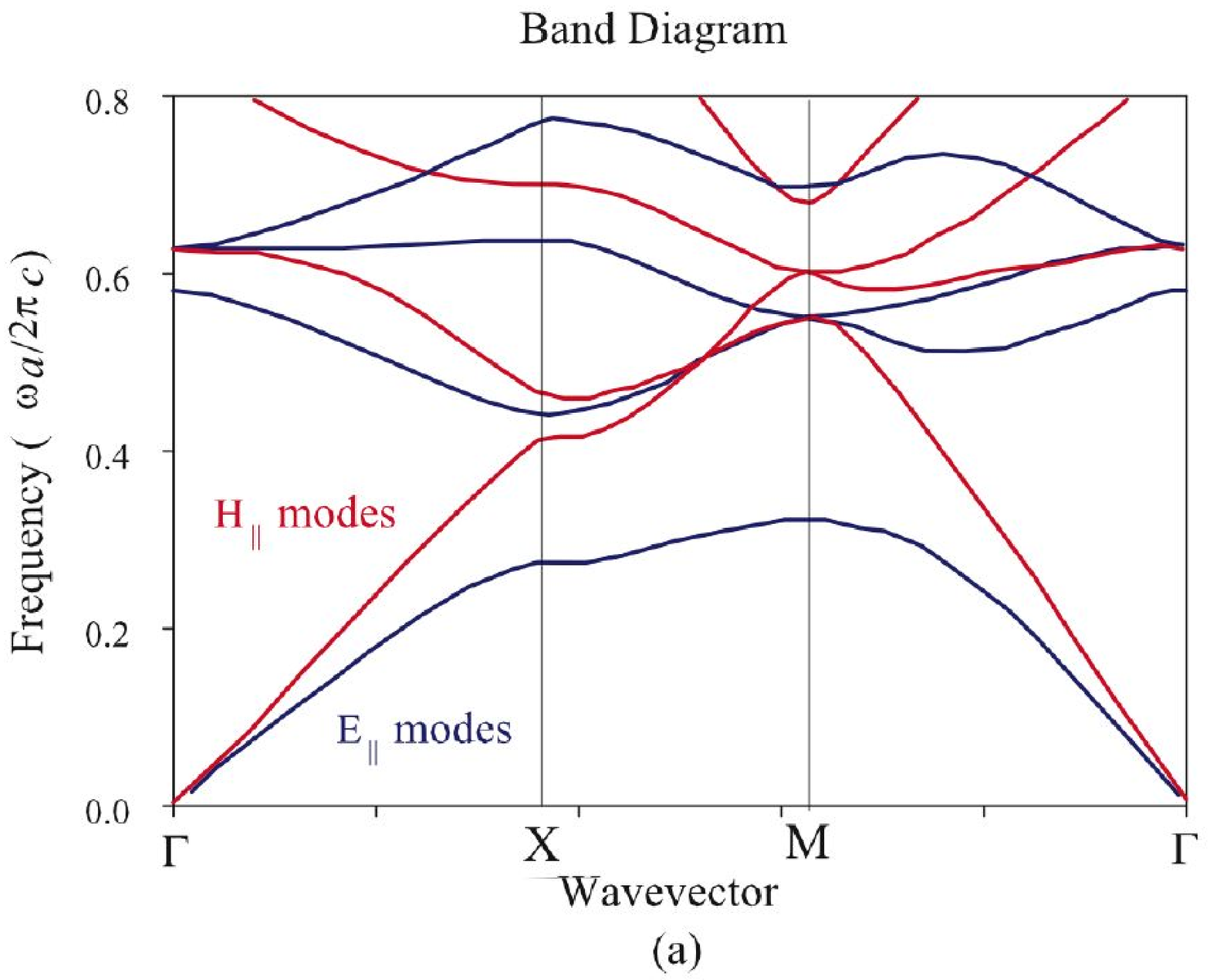}\includegraphics[width=6cm]{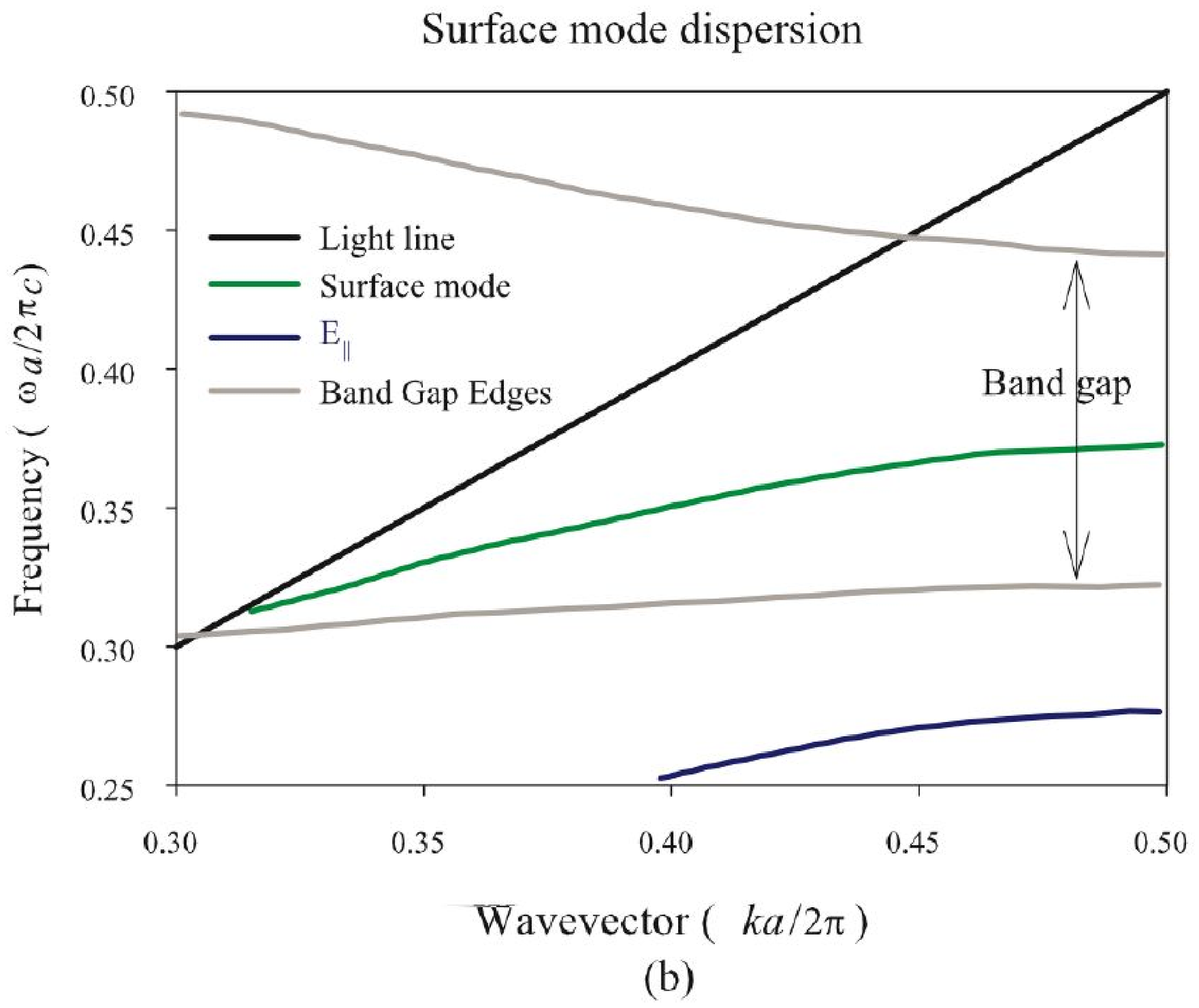}}
\caption{{\label{band}} (a) Band diagram of the bulk PC for an infinite 2D array of round rods with the electric vector parallel to the rods $(\textcolor{blue}{{\rm E_{||}}})$ (\textcolor{blue}{blue}) and the magnetic vector parallel to the rods $(\textcolor{red}{{\rm H_{||}}})$ (\textcolor{red}{red}). The vertical axis is normalized frequency $f = (\omega/c)(a/2 \pi)$. As shown by the \textcolor{blue}{blue} curves, an infinite PC of all round rods with the parameters given in Fig. \ref{pc} has a complete band gap in the frequency range $0.32 < f < 0.44 $ for ($\textcolor{blue}{{\rm E_{||}}}$) modes.  For $(\textcolor{red}{{\rm H_{||}}})$ modes, there is no complete bandgap. (b) These data are a subset of the $\Gamma$-X portion of (a) where the \textcolor{green}{green} curve shows the dispersion of ${\rm E_{||}}$ surface modes that propagate with wavevector $k$ parallel to the semi-circular cylinder interface at frequencies within the bandgap.  The black line is the light line and the gray lines show the band gap limits. Of particular interest in the analysis to follow is the \textcolor{green}{green} curve in (b) for the localized surface wave mode. We have chosen to truncate the top row of cylinders along their diameter, however varying the amount of truncation could be done and this would yield different dispersion curves for the surface mode. The data in (a) and (b) are reproduced from \cite{mit}. }
\end{figure}

We calculate the LDOS of a finite size 2D PC for the cases where the dielectric material is all linear or some rods are nonlinear. We will also investigate the response of the PC to an incident electric field. Of particular interest are localized surface wave modes along a PC interface. To model the 2D PC, we consider a finite-sized square array of dielectric cylinders as shown in Fig. \ref{pc}. We begin with the case where the PC consists of linear material and assume the electric vector is polarized parallel to the rods. Since the PC considered here is finite size, the possibility of defect and localized modes exist. The design parameters of the PC are such that a band gap exists and truncation of the topmost layer of rods permits localized surface wave modes to occur at frequencies within the band gap. The dispersion of the bulk PC and surface modes is shown in Fig. \ref{band}. 

In the case of linear media, the LDOS is independent of an incident field. In the case of non-linear material, we illuminate the PC with a plane wave and include non-linear material effects in a self-consistent manner. In this case, any change in the LDOS as calculated here, is examined. We neglect the effects of coupling to higher harmonic fields. 

\subsection{Dyson's equation and iterative solution}
The method of analysis involves iterative solution to Dyson's equation to calculate the Green function and the Lippman-Schwinger equation for the electric field.\cite{martin1,martin2,martin3} We assume harmonic fields that vary as ${\rm exp}(-i\omega t)$. Here we consider an especially simple case where the electric vector is polarized parallel to the axis of the dielectric cylinders. In this case, the Green dyadic ${{\bf G} = {\hat z}{\hat z} G_{zz}}$ effectively becomes a scalar that satisfies Dyson's equation
\begin{eqnarray} 
\label{e01} G_{zz}({\bf \rho}, {\bf \rho^{\prime}}) =  G_{zz}^0({\bf \rho}, {\bf \rho^{\prime}}) + k_0^2 \int_A d^2\rho^{\prime \prime}  G_{zz}^0({\bf \rho}, {\bf \rho^{\prime \prime}}) \hat\epsilon \left({\bf \rho^{\prime \prime}}\right)  G_{zz}({\bf \rho^{\prime \prime}}, {\bf \rho^{\prime}})
\end{eqnarray}
and the electric field ${\bf E} = {\hat z}E_z$ satisfies either of the following forms of the Lippmann-Schwinger equation,
\begin{eqnarray} 
\label{e02a}
E_z({\bf \rho}) = E^0_z({\bf \rho}) + k_0^2 \int_A d^2 \rho^{\prime} G_{zz}({\bf \rho}, {\bf \rho^{\prime}}) \hat\epsilon \left( {\bf \rho^{\prime}} \right) E^0_z({\bf \rho^{\prime}}) \\
\label{e02b}
E_z({\bf \rho}) = E^0_z({\bf \rho}) + k_0^2 \int_A d^2 \rho^{\prime} G^0_{zz}({\bf \rho}, {\bf \rho^{\prime}}) \hat\epsilon \left( {\bf \rho^{\prime}} \right) E_z({\bf \rho^{\prime}})
\end{eqnarray}
where $k_0 = \omega/c$ and ${\bf \rho} = (x,y)$. The integration in Eqs. (\ref{e01})-(\ref{e02b}) is only over the cross-sectional area of the cylinders. The $G_{zz}^0({\bf \rho}, {\bf \rho^{\prime}}) = \frac{i}{4}H_0\left( k_0|{\bf \rho} - {\bf \rho^{\prime}}|\right)$ is the Green function for free space in the absence of the PC structure and $G_{zz}$ is the Green function with the effects of the PC included. The $E^0_z({\bf \rho})$ is the incident electric field. Note that when ${\bf \rho} = {\bf \rho^{\prime}}$ in Eqs. (\ref{e01})-(\ref{e02b}) the singularity associated with the Green function  would normally require special treatment, but in the simple case treated here, the singularity is more easily handled.\cite{yaghjian} For simplicity, we omit all $z$ subscripts henceforth.

The PC structure is described by the $\hat\epsilon \left({\bf \rho}\right)$ that is non-zero only within the dielectric rods. Initially, we assume the rods are homogeneous with permittivity $\epsilon_c$ embedded in a background medium of unit permittivity. This yields
\begin{equation}
\label{e02.5}
\hat\epsilon \left({\bf \rho}\right)  =
 \epsilon_c - 1 
\end{equation}
when $\bf \rho$ is within any rod and zero otherwise. Numerical solution to Eqs. (\ref{e01}) and (\ref{e02b}) can be performed by direct inversion, but this would typically be very large in terms of memory requirements, especially in Eq. (\ref{e01}). Here we provide only a brief description of the iterative method. The computational space is divided into an array of grids where the $i$th grid has area $\Delta A$. The center of the $i$th and $j$th grid is located at ${\bf \rho}_i$ and ${\bf \rho}_j$. The space is initially empty and the PC is then constructed by adding one ``infinitesimal" $\hat\epsilon$ element at a time. When the $n$th element $\hat\epsilon_{k_n}$ is added we have
\begin{eqnarray} 
\label{e04}
G_{k_n j}^n = \left(I - M_{k_n}^{n-1} \hat\epsilon_{k_n} \right)^{-1} G^{n-1}_{k_n j} \qquad &i = k_n,&j \ne k_n \nonumber \\
G_{ij}^n = G_{ij}^{n-1} + {\hat A} G^{n-1}_{ik_n} \hat\epsilon_{k_n} G^n_{k_n j}  \qquad &i \ne k_n,& j \ne k_n\\
G_{ik_n}^n =  \left(I - M^n_{k_n} \hat\epsilon_{k_n} \right)^{-1} G_{ik_n}^{n-1} \qquad &i \ne k_n,& j = k_n \nonumber
\end{eqnarray}
where $k_0^2 \Delta A = {\hat A}$. To initiate the iteration algorithm, set $n=1$ and there is a single perturbation $\hat\epsilon_{k_1}$ at grid $k_1$. The $M_{k_1}^0$ term represents the integration term when $i=j=k_1$ as
\begin{eqnarray} 
\label{e05}
M_{k_1}^0 = k_0^2 \int_A \ G^{0}_{k_1 k_1} \ dA =  \frac{i \pi k_0 R H_1(k_0 R)}{2} - 1
\end{eqnarray}
where $R = \sqrt{\Delta A/\pi}$. In (\ref{e01}), the integral over the grid square has been approximated by integrating over an equivalent circular area of radius $R$. The electric field solution can be built up in a similar iterative fashion or calculated as 
\begin{eqnarray} 
\label{e06}
E_i = E^0_i + {\hat A}\sum_{n=1}^N G_{i k_n}^N \hat\epsilon_{k_n} E^0_{k_n} 
\end{eqnarray}
where $E^0_i$ is the incident field at the $i$th coordinate and $N$ is the number of perturbation elements $\hat\epsilon_{k_n}$ in the structure. We have given a very brief outline of the method since details\cite{martin1,martin2,martin3} are described elsewhere and the method is not the main focus of this work.

\subsection{Local density of states}
 Computation of the electric field and the Green dyadic yields much information, including the normalized LDOS, that is defined here as 
\begin{eqnarray} 
\label{e03}
{\rm LDOS}({\bf \rho}) = \frac{ \Im\left[ G \left({\bf \rho},{\bf \rho}\right)\right]} {\Im\left[ G^0 \left({\bf \rho},{\bf \rho}\right)\right]} \ ,
\end{eqnarray}
where $G^0$ is the free space Green function as given in Eq. (\ref{e01}), and following Eq. (\ref{e02b}), and $\Im$ indicates the imaginary part. In the case of linear media, the LDOS as calculated from Eq. (\ref{e03}) is only dependent on the material and structural parameters of the PC. If nonlinear material is introduced, then obviously an excitation field is required to activate the nonlinearity. In this case, we still use Eq. (\ref{e03}) to calculate the LDOS with the PC illuminated by an incident plane wave of unit amplitude and at a given angle of incidence. This is done self-consistently yielding changes in the permittivity as described in the next section. The resulting changes in permittivity yields different values for $G\left({\bf \rho},{\bf \rho}\right)$ and consequently the ${\rm LDOS}({\bf \rho})$. The modeling of the nonlinearity is discussed in the following section. 

In \cite{fussell}, the local density of states is written as ${\rm LDOS} = -{{2\omega}\over{\pi c^2}} \Im [G({\bf \rho},{\bf \rho})]$. Since in this work we consider the LDOS in Eq. (\ref{e03}) that is normalized to free space, we note that, $-\Im G^0({\bf \rho},{\bf \rho})= 0.25$. It follows that the normalized LDOS given in numerical results below can easily be compared to free space relative to $0.25$.

\subsection{Non-linear material}
Others have considered the effect of NLM on the infinite photonic crystal.\cite{tran,lousse,soljacic,bahl,mingaleev} Using the iterative procedure\cite{martin1, martin2, martin3} briefly described above, we calculate for a finite size PC the Green function, LDOS, and the response of the PC shown in Fig. \ref{pc} to an incident plane wave. To consider the addition of a NLM in the topmost layer of semi-circular rods, we take a self-consistent approach.\cite{lousse} Assuming a Kerr non-linearity we write

\begin{eqnarray}
\label{e10}
\hat\epsilon \left({\bf \rho}\right) = \Biggl\{
\begin{matrix}
{ &{\epsilon_c - 1 + \chi |E_c(\bf \rho)|^2 \ {: \ {\bf \rho} \ {\rm within \ any \ semicircular \ cylinder}}} \nonumber \\
&{\epsilon_c - 1 \ {: \ {\bf \rho} \ {\rm within \ any \ round \ cylinder}}} \nonumber \\
&{0 \ {: \ {\bf \rho} \ {\rm otherwise}}} } 
\end{matrix}
\end{eqnarray}
\begin{figure}
\centerline{\includegraphics[width=6cm]{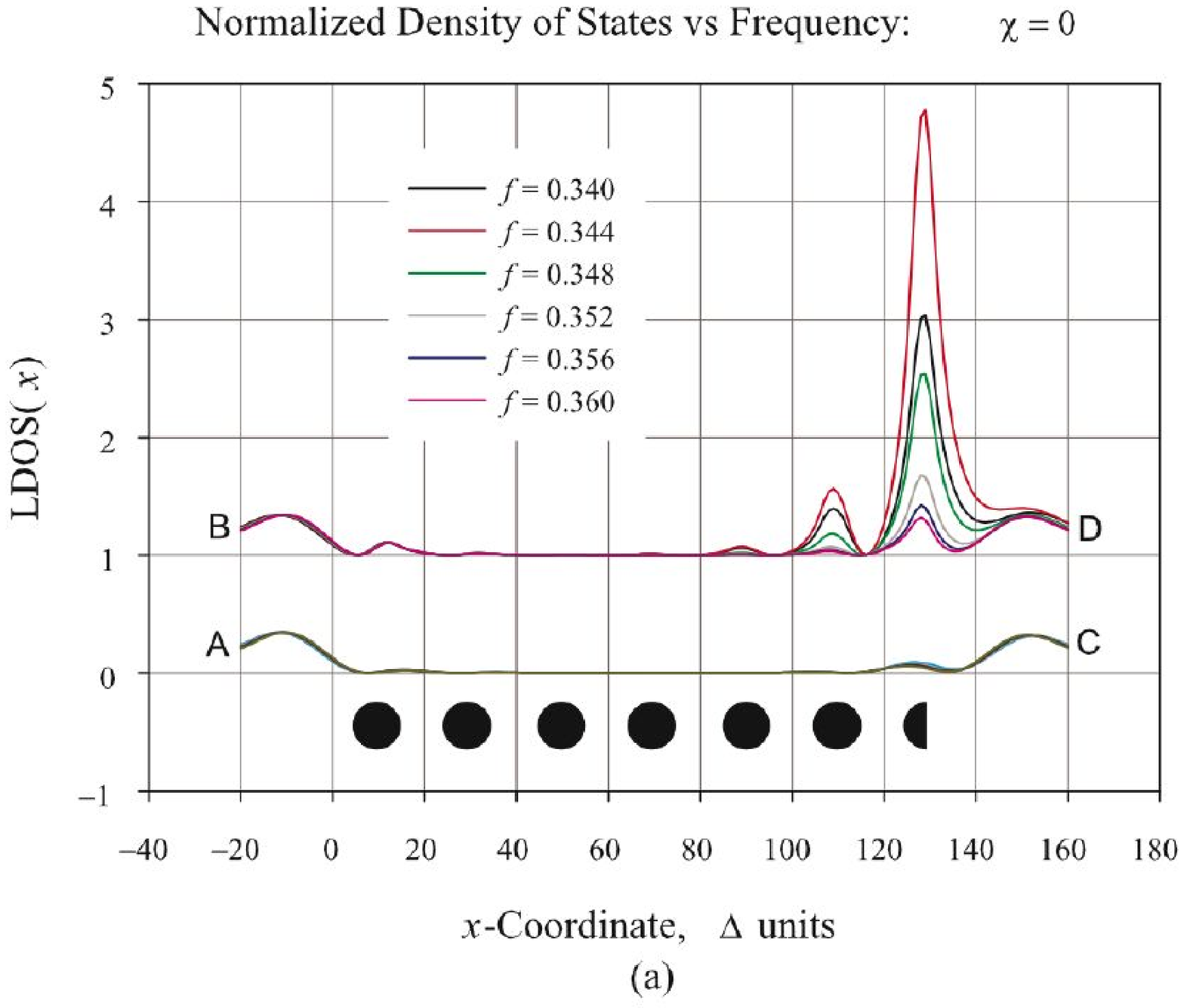} \includegraphics[width=6cm]{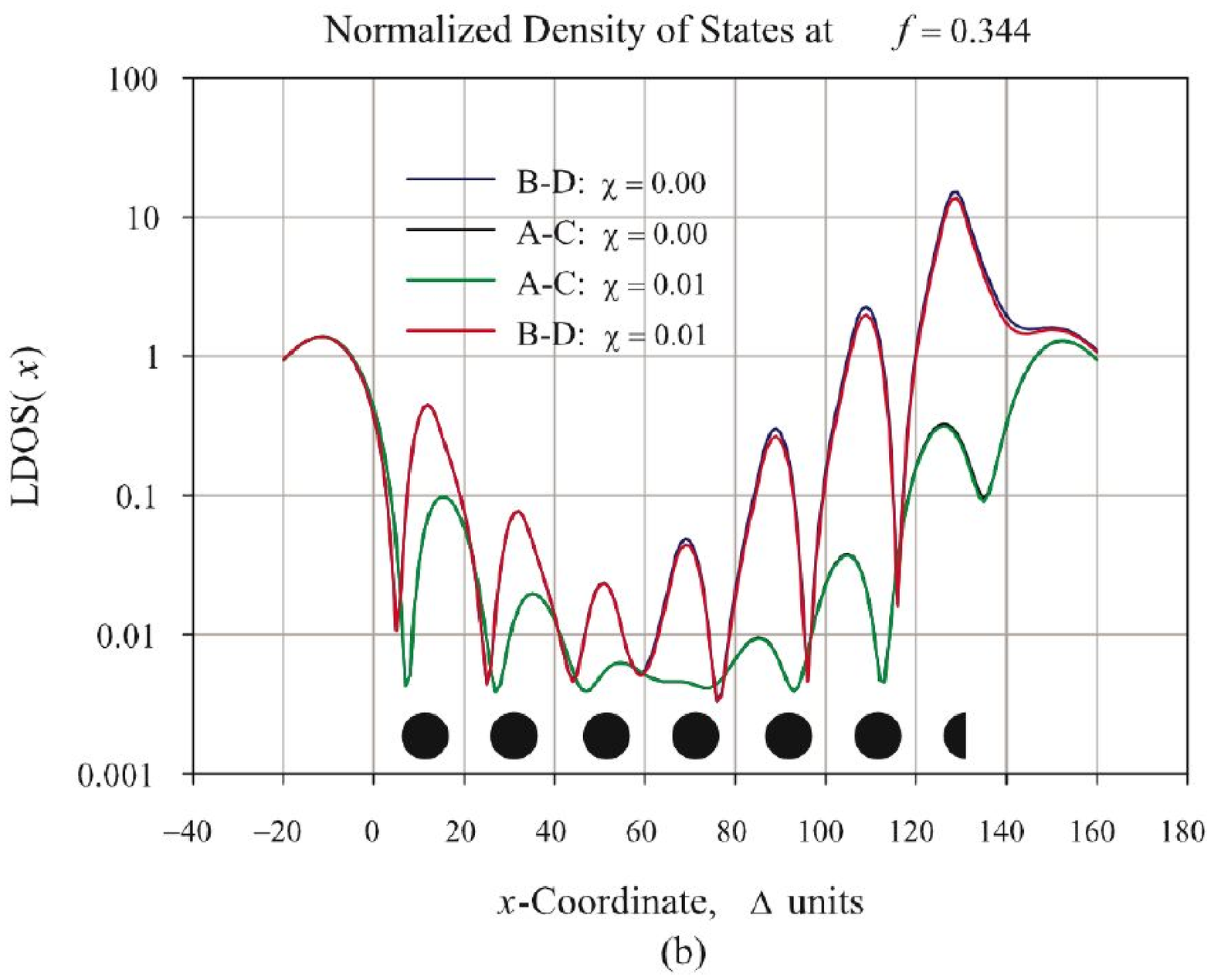}}
\caption{{\label{f02}}(a) Normalized LDOS({\it x}) vs position {\it x} within the PC for various frequencies with linear material and (b) Normalized LDOS({\it x}) at resonance frequency compared with non-linear material. In (a), and referring to Fig. 1, the LDOS({\it x}) is shown on a linear scale and is calculated along line AC (between rods) and BD (through rods) where the BD data has been displaced one unit upward for clarity. For reference, the large black circles in the lower part of the figures indicate the positions of the rods in relation to the LDOS. In (a), the dimensionless frequency parameter $f = (\omega/c)(a/2 \pi)$ is varied for $\omega$ within the band gap from $0.34$ to $0.36$ with a peak at $f = 0.344$. Along line BD, there is clear indication of localized states concentrated at the semi-circular rod interface ($x \approx 130\Delta$, where $\Delta$ is the spatial resolution) whereas no localization at the opposite circular rod interface ($x \approx 10\Delta$). Along line AC there is no comparable indication of localized states. It follows that the localized states exclusively reside within the semi-circular rods. This indicates the importance of defect parameters, such as altering the rod cross-section, in the surface mode dispersion. In (b), the resonant LDOS({\it x}) is plotted on a log scale along lines AC and BD for $f = 0.344$ with $\chi = 0$ and $0.01$. The $\chi=0$ curves are repeated from (a). To activate the NLM, the PC is illuminated by a plane wave of unit amplitude incident at $\theta = 45^o$. Along line AC, the green and black curves are superimposed. Along line BD, the red and blue curves are nearly superimposed with the blue curve slightly larger in value that indicates a slight reduction in LDOS with non-zero $\chi$. This reduction is not very clear on this log scale, however the log scale does show more clearly the LDOS({\it x}) throughout the PC. This indicates that the presence of non-linear material in the semi-circular rods does not significantly alter the LDOS. Finally, these numerical data are completely independent of an incident field.}
\end{figure}
\begin{figure}
\centerline{\includegraphics[width=6cm]{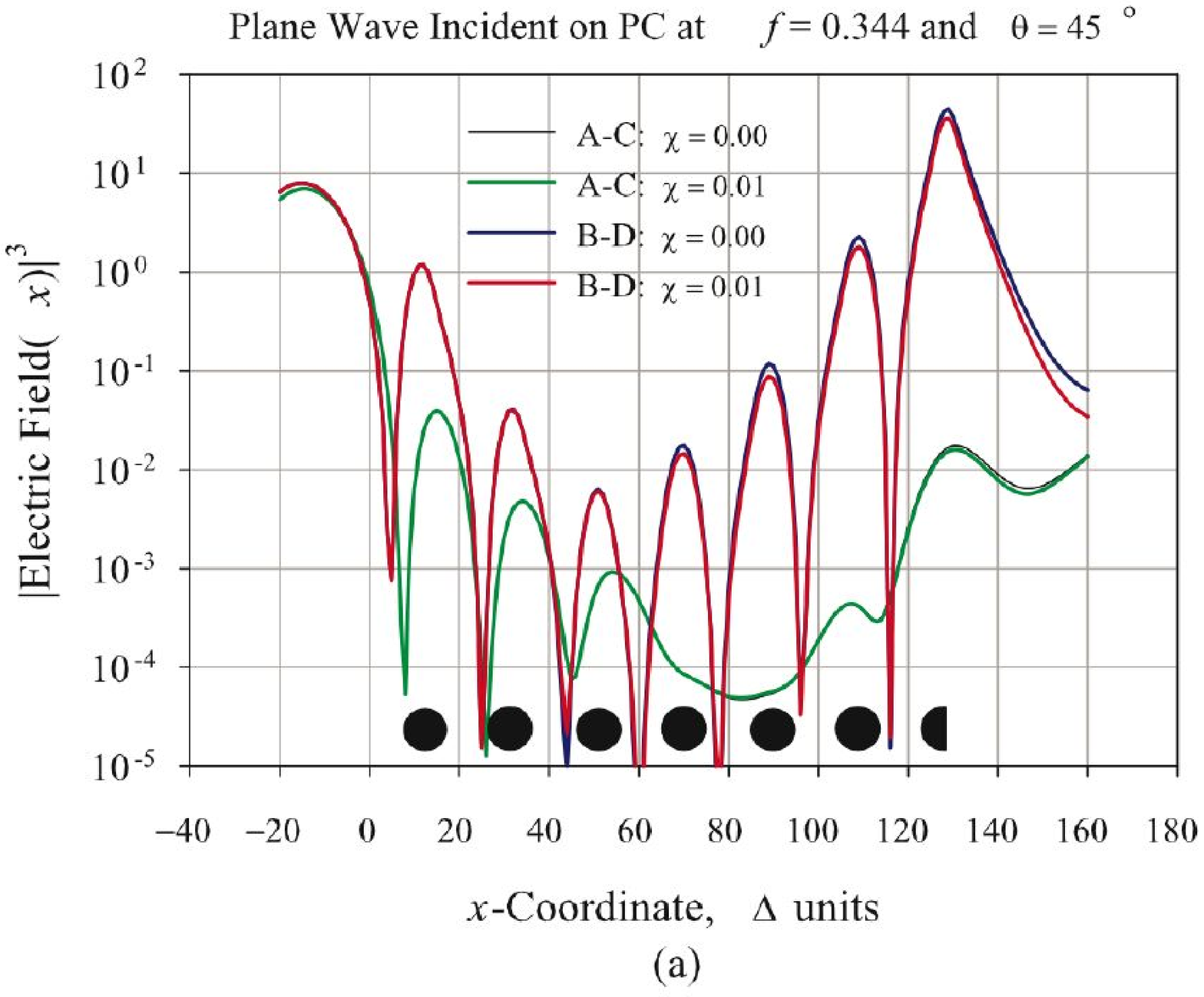} \includegraphics[width=6cm]{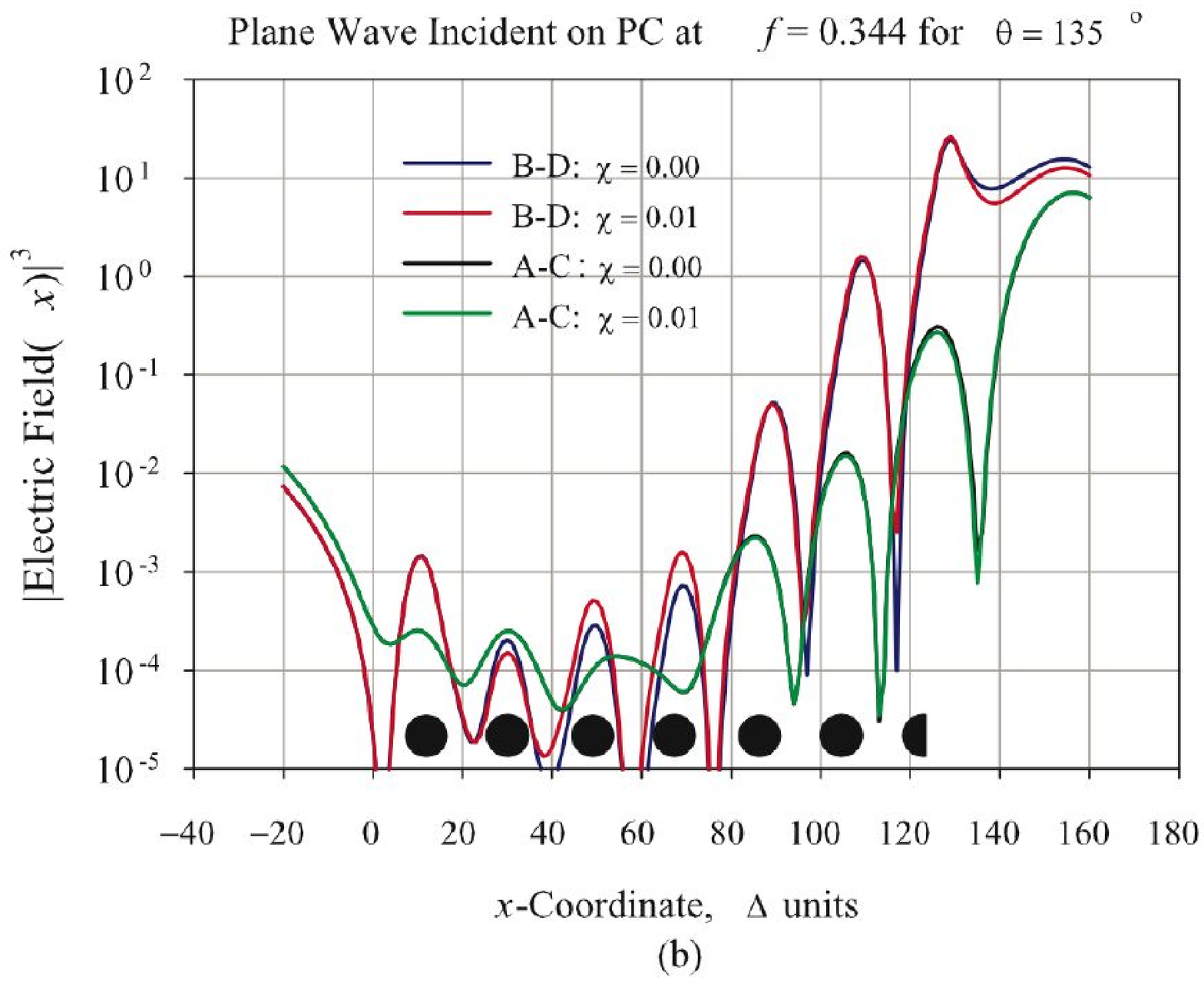}}
\caption{{\label{f03}}Plane wave at normalized frequency $f = 0.344$ incident on PC at angles of (a) $45^o$ and (b) $135^o$ (see Fig. \ref{pc}) for both linear and non-linear material. The {\it x}-coordinate is in $\Delta$ units and the electric field is plotted on a log scale as $|E|^3$ since this is relevant to a Kerr nonlinearity. For both angles of incidence, there is significant excitation of the surface defect mode as noted by the peaks around $x \approx 130\Delta$ and lines BD . In (a), there are incident and reflected fields in the $x \le 0$ PC interface region. In (b), there are again the incident and reflected fields for $x \ge 130\Delta$ in addition to the surface mode excitation field. Note also in (b) that the low field intensity for $x \approx 0$ region is because this is a near-field shadow region. It is seen that for a unit amplitude incident wave, the localized $|E|^3$ field is well in excess of an order of magnitude larger because of the surface mode excitation and the localization is precisely at the sites of the nonlinear material.}
\end{figure}
\begin{figure}
\centerline{\includegraphics[width=6cm]{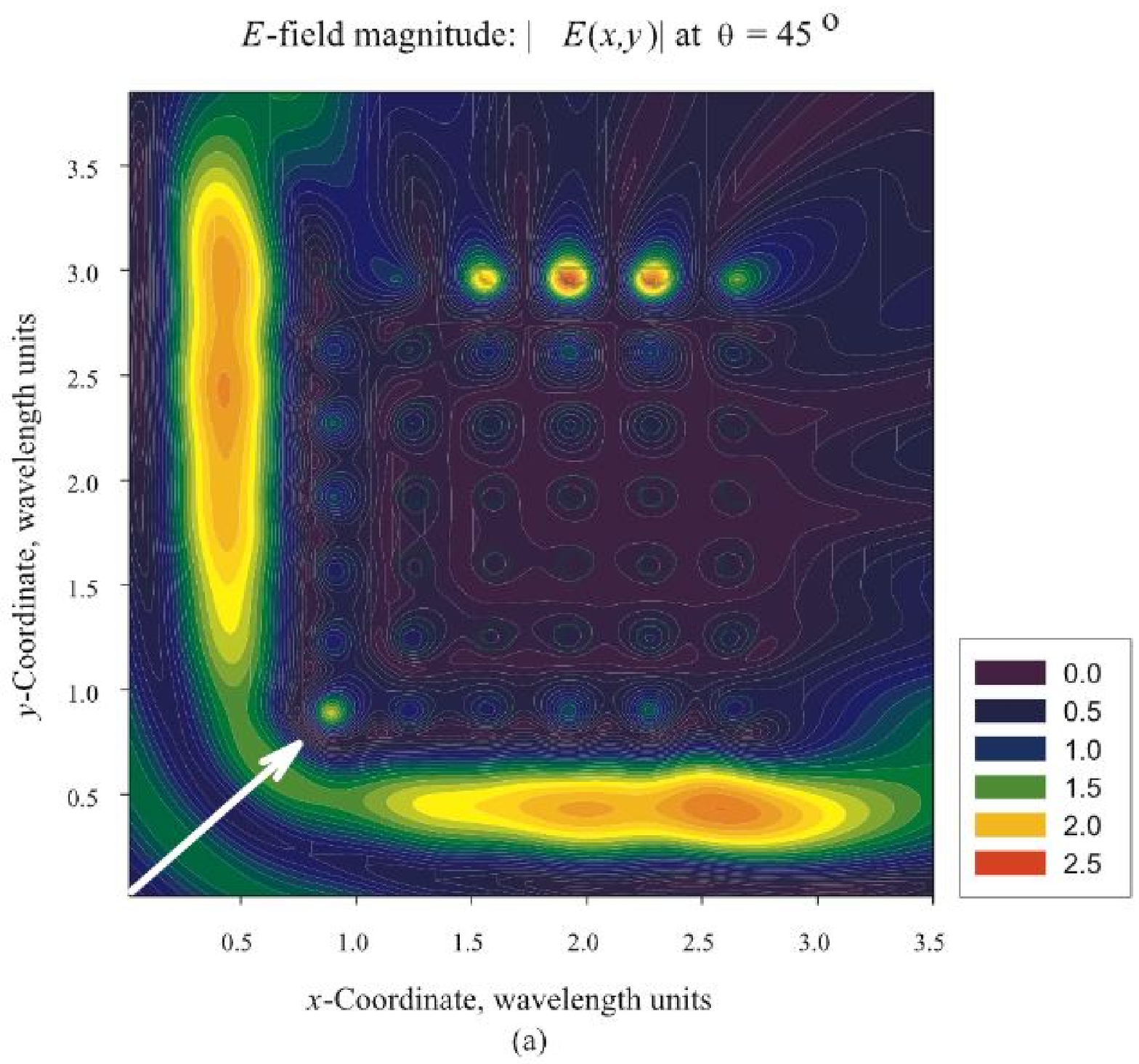} \includegraphics[width=6cm]{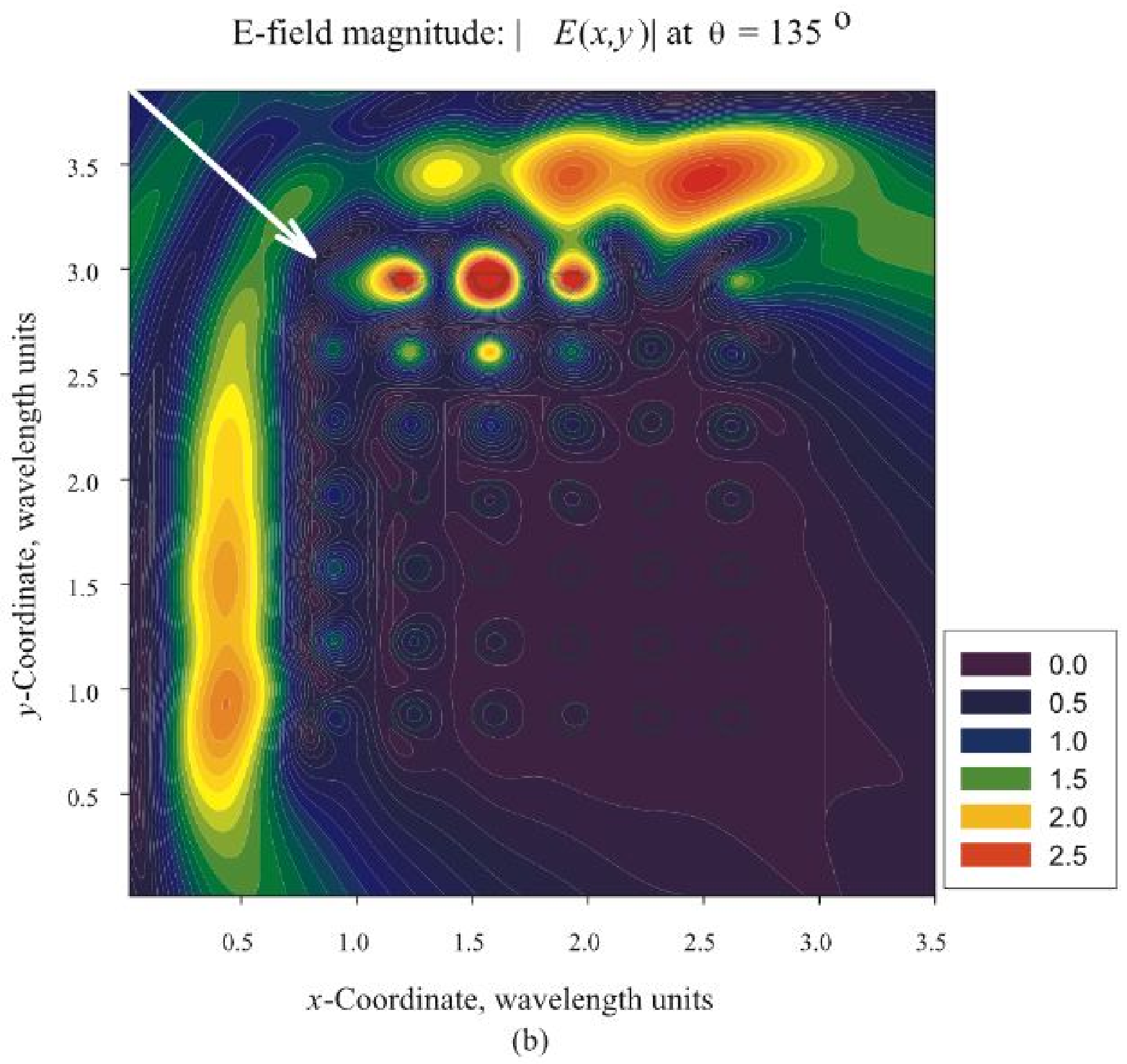}}
\caption{{\label{f04}}Contour plots of a plane wave at frequency $f = 0.344$ incident on PC at angles of (a) $45^o$ and (b) $135^o$ for linear material. The angles of incidence are indicated by the arrows. These plots show the electric field intensity throughout the PC and within 2 periods $a$ outside the PC., The line plots of Fig. \ref{f03} are a subset of these data. The {\it x} and {\it y}-coordinates are in $\lambda_0 = a/0.35$ units. Surface defect mode excitation is clearly seen along the truncated layer of rods. Note that the outline of the PC rod structure is superimposed on this contour plot.}
\end{figure}
Initially, the electric field distribution within the semi-circular rods, $E_c^0(\bf \rho)$, is obtained when $\hat\epsilon^0({\bf \rho}) =  \epsilon_c - 1$. Using $E_c^0(\bf \rho)$, the permittivity is then changed to $\hat\epsilon^1({\bf \rho}) = \epsilon_c - 1 + \chi |E_c^0({\bf \rho})|^2$. Using $\hat\epsilon^1({\bf \rho})$, a new Green function $G_{ij}$ and field distribution $E_c^1({\bf \rho})$ are calculated. This process is repeated until succeeding electric fields throughout the PC converge to a stable result. This yields not only the electric fields, but also any changes in the LDOS.

In testing for convergence of the self-consistent algorithm, we found that for $\chi \le 0.01$ approximately 10 iterations or less were needed to converge. If the discretized computational domain contains $M$ points $(x,y)$, we calculate the electric field $E^{(n)}(x,y)$ for all $M$ points. We assume convergence is satisfied when $ {{1}\over{M}}\sum_{x,y} \mid r^{(n)}(x,y) \mid \le 0.0001$ where $r^{(n)}(x,y) = [E^{(n)}(x,y)-E^{(n-1)}(x,y)] /E^{(n-1)}(x,y)$ and $n$ is the number of iterations. We did not do an extensive parametric evaluation of the self-consistent convergence conditions. We did note that when $\chi = 0.1$, the self-consistent algorithm failed to converge. In any case, we did not try to use experimentally obtained $\chi$ values, rather we note that $\chi \le 0.01$ is well within any realistic material values. 

\section{Numerical Results \label{results}}
The PC is modeled as a square array of dielectric rods with period $a$ and radius $r$ with physical parameters\cite{mit} chosen so that there is a complete band gap for the infinite crystal and this is shown in Fig. \ref{band}(a) and the defect mode surface wave dispersion is shown in Fig. \ref{band}(b). The computational space is divided into square grids of side dimension $\Delta$ and the PC period is $a = N \Delta$. We choose a frequency $\omega_0$ within the band gap where $(\omega_0/c)(a/2 \pi) = 0.350$. We set $N = 20$ and this yields a free space resolution of $\Delta = 0.0175 \lambda_0$ with $\omega_0/c = 2 \pi/\lambda_0$. The resolution in a cylinder is then degraded to $\Delta_c = \Delta \sqrt{8.9}$, and this is still adequate for accurate computation.

To check the accuracy of our results, we have compared calculation of the electric field as obtained from Eqs. (\ref{e02a}) and (\ref{e02b}). We note that for reasonable sized grid of $N$ points, Eq. (\ref{e02a}) can be solved for $E_z({\bf\rho})$ by direct inversion. Also, $E_z({\bf\rho})$ can be obtained from Eq. (\ref{e02b}) after first obtaining $G_{zz}^{N}({\bf\rho},{\bf\rho}^{\prime})$. We find that the electric fields computed by both methods are in excellent agreement and we conclude that the method of calculating $G_{zz}^{N}({\bf\rho},{\bf\rho}^{\prime})$ is also accurate.

Figure \ref{f02} shows the results of computations of the LDOS associated with the PC and the plane wave response of the PC. In Fig. \ref{f02}(a), the calculated LDOS is shown for several frequencies and a PC with all linear material. At the surface mode resonant frequency, and with the PC illuminated with a unit-amplitude plane wave at $\theta=45^o$, the computed effect of including NLM is shown in Fig. \ref{f02}(b).  The NLM coefficient is chosen as $\chi = 0.01$. The results of the LDOS calculations indicate a large concentration of states available at the truncated-rod interface. When the NLM coefficient $\chi$ is non-zero, there is noticeable decrease in the LDOS. This decrease is due to the change in the permittivity of the semi-circular rods and the surface wave mode is quite sensitive to any physical or material changes in the semi-circular interface.

In Fig. \ref{f03} we consider the response of the PC when illuminated by a plane wave at two different angles relative to the truncated rod interface. We plot the electric field as $|E(x,y)|^3$ since this is the magnitude relevant to Kerr nonlinearity. For both angles of incidence, there is strong excitation of the surface mode. It is seen that for $\chi = 0.0$ and $0.01$ for both angles of incidence $\theta = 45^o$ and $135^o$, the surface mode peaks are quite similar in both cases.  This is likely due to the fact that very little of the PC actually contains nonlinear material. Nevertheless, there are some differences in the $|E(x,y)|^3$ results when $\chi = 0.0$ and $0.01$. The addition of $\chi = 0.01$ causes a slight decrease in $|E(x,y)|^3$ and this is consistent with the LDOS data in Fig. \ref{f03}(b).

A more comprehensive view of the field profile is illustrated in Fig. \ref{f04}, which shows the corresponding contour plots relating to Fig. \ref{f03} for linear media. Here the field magnitude is plotted over the entire PC including two periods outside the PC boundary. The surface mode excitation at the semicircular rod sites is clearly seen. The field magnitude inside the PC is quite diminished and the location of the rods comprising the PC is shown by an overlay of an outline of the rods.

\section{Conclusions}
An analysis of the LDOS for surface wave modes on a truncated 2D photonic crystal interface has been accomplished using a Green function formalism. The PC is composed of circular rods (except for the semicircular interface) and the electric field is parallel to the rods. The results show that the presence of NLM does not significantly alter the LDOS as defined in this paper and this could be due to the NLM being a small fraction of the total volume of PC material. If the idea of using a 2D PC as a component in frequency conversion is viable, then it seems that additional volume of NLM would probably be needed. We have also neglected the fact that frequency conversion does occur and that there is an additional coupled equation to consider that would describe the generation of higher harmonic fields due to the Kerr nonlinearity. This may be a source of damping that could broaden the resonance associated with the surface wave mode. This could  reduce the conversion efficiency. 

\section*{Acknowledgments}
Support was provided by ONR Independent Laboratory In-house Research funds and DARPA MORPH funds. Valuable computing resources and assistance were available from the Arctic Region Supercomputing Center.
\end{document}